\documentclass[
    onecolumn,
    preprintnumbers,
    amsmath,
    amssymb,
    prd,
    showpacs
]{revtex4}

\usepackage{graphicx}%
\usepackage{amsmath}
\usepackage{amsthm}
\usepackage[normalem,normalbf]{ulem}
\usepackage{amssymb}
\usepackage{mathrsfs}
\usepackage{bbm}
\usepackage{bm}
\usepackage{enumerate}
\usepackage[all]{xy}

\newcommand{\del}{\partial}
\newcommand{\nn}{\nonumber}

\newcommand{\half}{\frac12}
\newcommand{\halfi}{\frac{i}{2}}
\newcommand{\beq}{\begin{equation}}
\newcommand{\eeq}{\end{equation}}
\newcommand{\beqa}{\begin{eqnarray}}
\newcommand{\eeqa}{\end{eqnarray}}
\newcommand{\bseq}{\begin{subequations}}
\newcommand{\eseq}{\end{subequations}}

\newcommand{\fr}{\frac}
\newcommand{\bg}{\textbf{g}}
\newcommand{\mn}{{\mu \nu}}

\newcommand{\bone}{\mathbbm{1}}
\newcommand{\bh}{\mathbbm{h}}
\newcommand{\bn}{\mathbbm{n}}
\newcommand{\bP}{\mathbb{P}}
\newcommand{\bM}{\mathbb{M} }
\newcommand{\bomega}{\boldsymbol{\Omega} }
\newcommand{\cA}{\mathcal{A} }

\newcommand{\cF}{\mathcal{F}}

\newcommand{\cP}{\mathcal{P} }

\newcommand{\cU}{\mathcal{U}}

\newcommand{\poin}{Poincar\'{e} }
\newcommand{\cFth}{\mathcal{F_\theta}}
\newcommand{\cFk}{\mathcal{F_\kappa}}

\begin{document}
%-------------------------%-------------------------------
\title{$\kappa$-deformed Spacetime From Twist}
%---------------------------------------------------------
\author{Jong-Geon Bu}
\email{bjgeon@yonsei.ac.kr}
%--------------------------------------------------------
\author{Hyeong-Chan Kim}
\email{hckim@phya.yonsei.ac.kr}
%---------------------------------------------------------
\author{Youngone Lee}
\email{youngone@phya.yonsei.ac.kr}
%--------------------------------------------------------
\author{Chang Hyon Vac}
\email{Shoutpeace@yonsei.ac.kr}
%----------------------------------------------------
\author{Jae Hyung Yee}%
\email{jhyee@phya.yonsei.ac.kr}
%---------------------------------------------------
\affiliation{Department of Physics, Yonsei University, Seoul, Korea.
}%
%----------------------------------------------------
\date{\today}%
%---------------------------------------------------
\bigskip
%----------------------------------------------------
\begin{abstract}
%-----------------------------------------------------
\bigskip

We twist the Hopf algebra of $igl(n,R)$ to obtain the
$\kappa$-deformed spacetime coordinates.  Coproducts of the twisted
Hopf algebras are explicitly given. The $\kappa$-deformed spacetime
obtained this way satisfies the same commutation relation as that of
the conventional $\kappa$-Minkowski spacetime, but its Hopf algebra
structure is different from the well known $\kappa$-deformed \poin algebra
in that it has larger symmetry algebra than the $\kappa$-Minkowski case.
There are some physical models which consider this symmetry
\cite{Percacci, Smolin, Floreanini}.
Incidentally, we obtain the canonical
($\theta$-deformed) non-commutative spacetime 
from canonically twisted $igl(n,R)$ Hopf algebra.

%-------------------------------------------------
\end{abstract}
%-------------------------------------------------
%------------------------------------------------
%------------------------------------------------
\maketitle
%-------------------------------------------------
%%%%

 \section{Introduction}

There have been extensive efforts to understand the gravity and quantum
physics in a unified viewpoint. These led to the developments in
many new directions of research in theoretical physics and mathematics.
To accommodate the quantum aspect
of spacetime, spacetime non-commutativity has been studied intensively
\cite{Seiberg,Gaume,chaichian0,Gomis,Bahns}. The majority of the
research in this direction focused on two types of noncommutative spacetimes
 \cite{Lukierski,Doplicher}, i.e., the canonical noncommutative
spacetime satisfying,
\begin{eqnarray}
\label{nc1}
\left[x^\mu, x^\nu\right] &=& i \theta^{\mn},
\eeqa

where $\theta^{\mu\nu}$ is a constant antisymmetric matrix,
and the commutation relation for the $\kappa$-Minkowski spacetime,
\begin{eqnarray}
\label{nc2}
\left[ x^\mu, x^\nu \right] &=& \frac{i}{\kappa} (a^\mu x^\nu-a^\nu
x^\mu),
\end{eqnarray}

where $\kappa$ is a parameter of mass dimension.
We call this $\kappa$-non-commutativity as
 time-like if $a^\mu a_\mu <0$, space-like if $a^\mu a_\mu
>0$,  and light-cone $\kappa$-commutation relation if $a^\mu a_\mu
=0$. In this paper, we consider time-like noncommutativity (i.e.
for $a^\mu\equiv(1,0,0,0)$).

The quantum field theories on canonical non-commutative spacetime
have many interesting features. By the Weyl-Moyal correspondence,
the theory can be thought as a theory on commutative spacetime with
Moyal product of the field variables \cite{Weyl}. The theories break
the classical symmetries (for example, Poincar\'{e} symmetry), and
may not satisfy unitarity, locality, and some other properties of
the corresponding commutative quantum field theory depending on the
structure of $\theta^{\mu \nu}$. The attempts to cure these
pathologies are still under
progress~\cite{Bahns1,Bahns2,BahnDo,Liao,Yee}.

%Connes and others studied extensively on the issues related with
%non-commutative geometry.
%There has been many researches
Extensive studies have been devoted to the quantum group theory
developed in the course of constructing the solutions of the
Quantum Yang-Baxter equations. A branch of quantum group theory, deformation
theory, was led by Drinfeld \cite{Drinfeld} and Jimbo \cite{Jimbo}.
Especially, Drinfeld \cite{Drinfeld} discovered a method of finding
one parameter solutions of the Quantum Yang-Baxter equation from simple Lie
algebra $\bg$, that is, he found one parameter family of solutions
of Hopf algebras $\cU_q(\bg)$ deformed from Hopf algebra of the universal enveloping
algebra $\cU(\bg)$ \cite{Baez}.

Recently, following Oeckl \cite{Oeckl}, Chaichian et.al. \cite{chaichian} and Wess \cite{Wess}
proposed a new kind of symmetry group which is deformed from the
classical Poincar\'{e} group. Especially, Chaichian
et.al.~\cite{chaichian} use the twist deformation of quantum group
theory to interpret the symmetry of the canonical noncommutative
field theory as twisted Poincar\'{e} symmetry. There have been some
attempts to apply the idea to the field theory with different
symmetries, for example, to theories with the $\Theta-$Poincar\'{e}
symmetry \cite {Gonera}, conformal symmetry
\cite{Matlock},\cite{Lizzi}, super conformal symmetry
\cite{Choonkyu}, super symmetry\cite{Saemann,Kobayashi}, Galilean symmetry
\cite{Sunandan}, Galileo Schr$\ddot{o}$dinger symmetry
\cite{Banerjee}, translational symmetry of $R^d$ \cite{Oeckl}, gauge
symmetry \cite{Vassil},\cite{Wess0}, diffeomorphic symmetry
\cite{Archil,Aschieri,Aschieri2}, and fuzzy diffeomorphism
\cite{Saemann2}.  The virtue of these twists is
that the irreducible representation of the twisted group does not
change from that of the original untwisted group, and moreover, the
Casimir operators remain the same. And there have been some studies
in constructing consistent quantization formalism of field theory
with this twisted symmetry group~\cite{chaichian2,Bala,Zahn,Bu}.

In the same reasoning,
if one finds a twist that gives $\kappa$-deformed
commutation relation
(especially time-like $\kappa$-deformed noncommutativity)
between coordinates from Poincar\'{e}  Hopf
algebra, it would be very useful in constructing quantum field
theory in $\kappa$-deformed spacetime since we can use the
irreducible representations of Poincar\'{e}  algebra for the
$\kappa$-noncommutative quantum field theory.
There has been some attempts
to obtain the $\kappa$-commutation relation of the coordinate system,
Eq.~(\ref{nc2}), from twisting the \poin Hopf algebra
\cite{Lukierski2},\cite{LukiWoro},\cite{LukiWoro2}.
In \cite{Lukierski2}, Lukierski et.al. have argued that
one can only get a light-cone
$\kappa$-deformation from the Poincar\'{e} Hopf algebra.
Hence, as far as we know, there is no twist of the Poincar\'{e}
algebra which gives a time-like $\kappa$-deformed coordinate
commutation relation.

This is our motivation to seek for the twist
that gives $\kappa$-deformed commutation relation from different
Hopf algebra which is larger than the Poincar\'{e} algebra.
The paper is organized as follows. In section \ref{Hopf}, we
briefly review the Hopf algebra and twist deformation. We present
two types of abelian twist elements,
leading to the two twisted spacetime coordinate systems
($\kappa$-deformed and $\theta$-deformed spacetime)
 in section \ref{igl}.
 We show that affine Hopf algebra, $igl(n,R)$, gives the
correct commuation relation by twisting.
In section \ref{discussion},
we discuss some aspects of the symmetry algebra
and $\kappa$-deformed spacetime induced from twisting $igl(n,R)$
with some physical examples.

\section{Brief Summary of twisting Hopf algebra}
\label{Hopf}

%\subsection{ Basic Hopf algebra}

For any Lie algebra $\bg$, we have a unique universal enveloping
algebra $\cU(\bg)$ which preserves the central property of the Lie
algebra (Lie commutator relations) in terms of unital associative
algebra \cite{Majid}. This $\cU(\bg)$ becomes a Hopf algebra if it
is endowed with a co-algebra structure. For $Y\in \cU(\bg)$,
$\cU(\bg)$ becomes a Hopf algebra if we define
 \begin{eqnarray}
\label{coproduct}
\Delta : \cU(\bg)\rightarrow \cU(\bg)\otimes \cU(\bg),
&&\Delta Y = Y\otimes 1+ 1\otimes Y, \nn\\
\epsilon(Y) = 0, &&S(Y)= -Y,
\end{eqnarray}
where $\Delta Y$ is a coproduct of $Y$, $\epsilon(Y)$ is a counit,
and $S(Y)$ is a coinverse (antipode) of $Y$. In other words, the set
$\{\cU(\bg),\cdot,\Delta,\epsilon,S\}$ constitutes a Hopf algebra.

$Y$ acts on the module algebra $\cA$ and on the tensor algebra of
$\cA$, and the action satisfies the relation (hereafter we use
Sweedler's notation $\Delta Y =\sum Y_{(1)}\otimes Y_{(2)} $
\cite{Majid})
\begin{equation}
Y\rhd (\phi \cdot \psi) = \sum (Y_{(1)}\rhd \phi) \cdot (Y_{(2)}\rhd \psi),
\label{action}
\end{equation}
where $\phi, \psi \in \cA$, the symbol $\cdot$ is a multiplication in the algebra $\cA$,
and the symbol $\rhd$ denotes the action of the Lie generators $Y\in U(\bg)$ on the module algebra $\cA$.

%\subsection{Deformation of Hopf algebra and module algebra}

We have a new (twisted) Hopf algebra $\{\cU_\cF(\bg),\cdot,\Delta_\cF,\epsilon_\cF,S_\cF\}$
from the original $\{\cU(\bg),\cdot,\Delta,\epsilon,S\}$ if there exists a twist element
$\cF \in \cU(\bg)\otimes \cU(\bg)$, which satisfies the relations
\begin{eqnarray}
\label{2cocycle}
(\mathcal{F}\otimes 1)\cdot (\Delta\otimes \mbox{id})\mathcal{F}
&=&(1\otimes \mathcal{F})\cdot  (\mbox{id}\otimes \Delta)\cF,\\
(\epsilon \otimes \mbox{id}) \cF=&1&=(\mbox{id}\otimes \epsilon)\cF
\label{counital}.
\end{eqnarray}
This relations are called counital 2-cocycle condition. The relation between the two Hopf algebras,
$\cU_\cF(\bg)$ and $\cU(\bg)$, is
\begin{eqnarray}
\Delta_\cF Y = \cF \cdot \Delta Y \cdot \cF^{-1}
&,&
\epsilon_\cF (Y) = \epsilon (Y), \nn\\
S_\cF (Y) =  u \cdot S(Y)\cdot u^{-1}
&,&u = \sum \cF_{(1)}\cdot S(\cF_{(2)}).
\label{tHopf}
\end{eqnarray}

If $\mathcal{A}$ is an algebra on which $\cU(\bg)$ acts covariantly in the sense of Eq.~(\ref{action}),
then
\begin{equation}
\label{tproduct}
\phi * \psi = \cdot ~[\cF^{-1}\rhd (\phi\otimes\psi)],
\end{equation}
for all $\phi, \psi \in \cA$, defines a new associative algebra $\cA_\cF$.
In constructing new Hopf algebra and getting a twisted module algebra, Eq.(\ref{2cocycle}) is  crucial for the associativity of the twisted module algebra.

%\subsection{ Applications to the physics}

This construction of twisted Hopf algebra has great advantages when
it is applied to physical problems whose symmetry group and the
irreducible representations are known, since we can use the same
irreducible representations and Casimir operators in the twisted
theory.
%As in the Chaichian et.al.¡¯s work,
%the relation Eq.(\ref{tproduct}) can be used to match the twisted theory to the
%field theory in non-commutative spacetime.

 \section{$\kappa$-deformed commutation relations from twisting $igl(n,R)$}
\label{igl}

In this paper
we focus on twisting the Hopf algebra of $igl(n,R)$
as a symmetry algebra.
In this section we presents two abelian twists which result in two
non-commutative coordinate systems,
Eq~(\ref{nc1}) and Eq~(\ref{nc2}).

We use the commutation relation of the Lie algebra, $\bg=igl(n,R)$,
\begin{eqnarray}
\label{Weylbasis}
\left[P_\mu, P_\nu\right] =0
&&
\left[ M^\mu_{~\nu}, P_\sigma \right] =
 i\delta^\mu_{~\sigma}\cdot P_\nu \nn \\
\left[M^\mu_{~\nu} ,M^\lambda_{~\tau}\right] &=&
i\left(\delta^\mu_{~\tau}\cdot M^\lambda_{~\nu}
-\delta^\lambda_{~\nu}\cdot M^\mu_{~\tau}\right).
\end{eqnarray}
where $P_\mu$ can be interpreted as generators of
the translation in the $x^\mu$-direction
and $ M^\mu_{~\nu}$ can be generators of rotations,
dilations and contractions.

In coordinate space, the generators are represented as
\begin{eqnarray}
\label{P:M}
P_\mu \rightarrow -i \partial_\mu, ~~ M^\mu_{~\nu}= -i x^\mu\partial_\nu .
\end{eqnarray}
(note that generators $ M^\mu_{~\nu}$ are different from
those of the \poin algebra.)

\subsection{$\kappa$ deformed non-commutativity}
\label{ksection}

The $\kappa$-deformation is generated by the twist element $\cFk$,
\begin{eqnarray}
\label{kelement}
{\cal F_\kappa}=
\exp\left[\frac{i}{2\kappa}\left( E\otimes D-D\otimes E\right)\right],
\end{eqnarray}
where
%we use the definition of the combination of generators
\begin{eqnarray}
\label{DE}
 D=h^\mu_{~\nu} M^\mu_{~\nu},~~~~
 E= n^\mu P_\mu .
\end{eqnarray}
$h^\mu_{~\nu}$ and $n^\mu$ are defined as $h^\mu_{~\nu}\equiv
\delta^\mu_{~\nu}-\delta^\mu_{~0}\delta^0_{~\nu},(n^\mu)=(1,0,0,0)$, that is, $D=\sum M^k_{~k},~ E=P_0$.
%here $\delta^\alpha_{~\beta}$ denotes the kronecker delta.
 These generators ($E,D$) commute with each other,
\begin{eqnarray}
\label{getED} ~[E, D] = -i n^\nu h^\mu_{~\nu} P_\mu=0 .
\end{eqnarray}
Hence we confirm that $\cF_\kappa$ is an abelian twist element.

%since $[E,D]=0$.
With this twist element Eq.~(\ref{kelement}), we twist the Hopf
algebra of $\cU(\bg\backsimeq igl(n,R))$ to get $\cU_\kappa(\bg)$
as in section \ref{Hopf}. From the fact that $[E\otimes D,D\otimes
E]= 0$, we can rewrite $\cF_\kappa$ as \beqa \cF_\kappa\equiv
\exp\left[\frac{i}{2\kappa}\left( E\otimes D\right)\right]
\exp\left[-\frac{i}{2\kappa}\left( D\otimes E\right)\right], \eeqa
which greatly simplifies the calculation of co-product of twisted
Hopf algebra. From the commutation relations~(\ref{Weylbasis}) we
have
\begin{eqnarray}
\label{D:M}
~[D, M^\mu_{~\nu}]=i(\bomega M)^\mu_{~\nu}, ~~~~
    \bomega^{\mu\nu}_{~~\rho\sigma}=
    \delta^\mu_{~\rho} h^\nu_{~\sigma}-h^\mu_{~\rho}\delta^\nu_{~\sigma}
    \equiv (\bone\otimes \bh-\bh\otimes \bone)^{\mu\nu}_{~~\rho\sigma}.
\end{eqnarray}
%Note that
From
%the simplified relation $\bomega\equiv \bone\otimes \bh- \bh\otimes \bone$ and
$\bh^2=\bh$, we get the relation
\begin{eqnarray}
\label{Om} \bomega^3=\bomega .
\end{eqnarray}

With the commutation relations~(\ref{Weylbasis}), we obtain the explicit forms of the coproduct $\Delta(Y)$,
\begin{eqnarray}
\label{DkGl}
\Delta_\kappa(\bP)|_\mu &=&\left. \left\{
    e^{\bh\otimes E/(2\kappa)}( \bP\otimes 1)
    + e^{-E/(2\kappa)\otimes \bh}(1\otimes \bP)\right\}\right|_\mu, \nn\\
\left. \Delta_\kappa (\bM)\right|^\mu_{~~\nu}&=&
 e^{\bomega\otimes E/(2\kappa)}(\bM\otimes 1)
 + e^{-\bomega\otimes E/(2\kappa)}(1\otimes \bM)\nn\\
&&+\frac{1}{2\kappa}\bn\cdot[e^{\bh\otimes E/(2\kappa)}(\bP\otimes D)
 -e^{-E/(2\kappa)\otimes \bh}(D\otimes \bP)] |^\mu_{~~\nu}.
\end{eqnarray}
In this calculation, we use the well-known operator relation,
$\displaystyle \mbox{Ad} e^B C=\sum_{n=0}^{\infty} \frac{(\mbox{Ad}
B)^n}{n!} C$, with $(\mbox{Ad} B)C=[B,C]$, and the relation
\begin{eqnarray}
 \label{DkY}
\Delta_\kappa(Y)
&=& \mbox{Ad} e^{\frac{i}{2\kappa}\left(E\otimes D-D\otimes E\right)}
 \Delta(Y)\nn\\
&=&e^{\frac{i}{2\kappa}\left(E\otimes D-D\otimes E\right)}
 \Delta(Y)e^{-\frac{i}{2\kappa}\left(E\otimes D-D\otimes E\right)}\nn\\
&=&e^{\frac{i}{2\kappa}\left(E\otimes D \right)}e^{-\frac{i}{2\kappa}\left(D\otimes E\right)}
(Y\otimes 1)
e^{\frac{i}{2\kappa}\left(D\otimes E\right)}e^{-\frac{i}{2\kappa}\left(E\otimes D \right)}\nn\\
&&~~~~~+e^{-\frac{i}{2\kappa}\left(D\otimes E\right)}e^{\frac{i}{2\kappa}\left(E\otimes D \right)}
(1\otimes Y)
e^{-\frac{i}{2\kappa}\left(E\otimes D \right)}e^{\frac{i}{2\kappa}\left(D\otimes E\right)}\nn\\
&=&\mbox{Ad} e^{\frac{i}{2\kappa}(E\otimes D)}\mbox{Ad} e^{-\frac{i}{2\kappa}(D\otimes E)}(Y\otimes 1)
+\mbox{Ad} e^{-\frac{i}{2\kappa}(D\otimes E)}\mbox{Ad} e^{\frac{i}{2\kappa}(E\otimes D)}(1\otimes Y).
\end{eqnarray}
Note that, from $\bh^2=\bh$ and $\bomega^3 =\bomega$, we have relations
\begin{eqnarray}
 \label{hOmega}
e^{\bh\otimes E/(2\kappa)}&=& 1\otimes 1+
\bh\otimes\left(e^{E/{2\kappa}}-1\right),\\
e^{\bomega\otimes  E/{2\kappa}}&=& 1\otimes 1+
\bomega\otimes \sinh\left(\fr{E}{2\kappa}\right)
+\bomega^2\otimes
\left[\cosh\left(\fr{E}{2\kappa}\right)-1\right] .
\end{eqnarray}
The algebra acts on the spacetime coordinates $x^\mu$ with commutative multiplication:
\begin{eqnarray}
\label{commul}
m(f(x)\otimes g(x)) := f(x) g(x) .
\end{eqnarray}

When twisting $\cU(\cP)$, one has to redefine the multiplication as in Eq.~(\ref{tproduct}),
while retaining the action of the generators of the Hopf algebra on the coordinates as in~(\ref{P:M}):
\begin{eqnarray}
 \label{kmult}
m_\kappa(f(x)\otimes g(x))&:=& f(x)\ast g(x)=m\left[\cFk^{-1}
    (f(x)\otimes g(x))\right].
\end{eqnarray}
It is represented as:
\begin{eqnarray}
 \label{kmoyal}
( f\ast g)(x)&:=&
\left. \exp \left[ \frac{i}{2\kappa}(\fr{\del}{\del x_0} y^k\fr{\del}{\del y_k}
-x^k \fr{\del}{\del x_k}\fr{\del}{\del y_0}) \right] 
f(x)g(y)\right|_{x=y}.
\end{eqnarray}

Since $P_\alpha = -i\del_\alpha$ in this representation, the commutation relations between spacetime coordinates
are deduced from this $*-$product:
\begin{eqnarray}
\label{kspace}
x^\mu *  x^\nu &=&
\cdot \left[e^{\frac{i}{2\kappa}(\del_0\otimes x^k\del_k- x^k\del_k\otimes\partial_0)}
\triangleright (x^\mu \otimes x^\nu) \right] \nn\\
 &=&
x^\mu\cdot x^\nu
+ \frac{i}{2\kappa}\left ( \delta^\mu_{~0}\delta^\nu_{~k}x^k -x^k\delta^\mu_{~k}\delta^\nu_{~0} \right), \nn \\
 \Longrightarrow [x^0, x^k]_*&=& \frac{i}{\kappa}x^k, ~~~~~~~~~~ [x_i,x_j]=0,
% ,  ~~~~~0~~\text{otherwise},
\end{eqnarray}
which corresponds to those of the time-like
$\kappa$-deformed spacetime.

The case of the tachyonic ($a_\mu
a^\mu=1$) and light-cone ($a_\mu a^\mu=-1$) $\kappa$-deformation
is obtained in Lukierski et.al.'s work \cite{LukiWoro}.
It should be noted that the twisted Hopf algebra in this section is different from
that of the conventional $\kappa$-Minkowski algebra, which is a
deformed Poincare algebra,
in that it has different co-algebra structure
from a bigger symmetry.

%supplemented with other commutation relations $[x_i,x_j]=0$.

\subsection{$\theta$-deformed non-commutativity}
\label{thsection}

Since affine algebra contains \poin algebra,
there is also a twist which has
same form as that of the canonical non-commutativity case
\cite{chaichian}.
We use the same twist element
given by
 \begin{eqnarray}
 \cF_\theta = \exp\left(\halfi \theta^{\alpha\beta} P_\alpha\otimes P_\beta\right).
\label{thelement}
\end{eqnarray}

This twist element satisfies the 2-cocycle condition, Eq.(\ref{2cocycle}).
Following the same procedures as in subsection (\ref{ksection})
we get the twisted Hopf algebra given by the coproducts,
\begin{eqnarray}
\label{DthGl}
 \Delta_\theta(P_\mu) &=& P_\mu\otimes 1+ 1\otimes P_\mu, \\
\Delta_\theta (M^\mu_{~\nu})&=& M^\mu_{~\nu}\otimes 1 + 1\otimes
M^\mu_{~\nu} +\half \theta^{\alpha\beta} \cdot[\delta^\mu_{~\alpha}
P_\nu\otimes P_\beta + \delta^\mu_{~\beta}P_\alpha\otimes P_\nu ].
\nonumber
\end{eqnarray}
Here also note that generators $ M^\mu_{~\nu}$ are different from
those of the \poin algebra.

Similarly as in the $\kappa$-deformed case, we have used
\begin{eqnarray}
 \label{DthY}
\Delta_\theta(Y)
&=& e^{\halfi \theta^{\alpha\beta} P_\alpha\otimes P_\beta}
\left(Y\otimes 1 + 1\otimes Y\right)
 e^{-\halfi \theta^{\alpha\beta} P_\alpha\otimes P_\beta}
 \nn\\
&=&\mbox{Ad} \exp\left(\halfi \theta^{\alpha\beta} P_\alpha\otimes P_\beta\right) \Delta(Y),
\end{eqnarray}
and
%from the commutation relations of Eq.(\ref{Weylbasis}), we have
\beqa
 \left[\mbox{Ad} (P_\alpha\otimes P_\beta)\right]^n
(M^\mu_{~\nu}\otimes 1) &=&
-i\delta^n_{~1}\cdot\delta^\mu_{~\alpha}(P_\nu\otimes P_\beta),\nn\\
\left[\mbox{Ad} (P_\alpha\otimes P_\beta)\right]^n (1\otimes
M^\mu_{~\nu}) &=&
 -i\delta^n_{~1}\cdot\delta^\mu_{~\beta}(P_\alpha\otimes
P_\nu),
 \eeqa
for $n \geq 1$.

When $\phi, \psi \in \cA_\theta$ are the functions  of the same spacetime coordinate $x^\mu$,
the product $*$ becomes the well known Moyal product. Since $P_\alpha = -i\del_\alpha$ in this
representation, the commutation relations between spacetime coordinates are
%is deduced from this $*$-product:
\begin{eqnarray}
\label{thcoor}
 x^\mu * x^\nu &=&
 \cdot ~\left[\exp\left(\halfi \theta^{\alpha\beta} \del_\alpha\otimes \del_\beta\right)
 \rhd (x^\mu\otimes x^\nu)\right]\nn\\
 &=& x^\mu\cdot x^\nu + \halfi \theta^{\mn},
\eeqa

which leads to the commutation relation
\begin{eqnarray}
 [x^\mu,x^\nu]_* &=& i \theta^\mn.
\end{eqnarray}

This is the same commutation relation of coordinates as those in the
canonical noncommutative spacetime. This twist is different from the
conventional twist \cite{chaichian}, in that the relevant group is
different.
The twist in the work of Wess\cite{Wess} and Chaichian, et
al.\cite{chaichian} is that of the Poincar\'{e}  Hopf algebra. Since
we use bigger symmetry algebra, $igl(n,R)$, than the Poincar\'{e}
algebra, $iso(n,R)$, only antisymmetric part of our twisted
coproduct of  generators $M^\mu_{~\nu}$ corresponds to those of
\cite{Wess} and \cite{chaichian}. We have more components
(coproducts of the symmetric part of the generators $M^\mu_{~\nu}$)
in the coproduct sector.
\section{Discussion}
\label{discussion}

In this paper we obtained
time-like $\kappa$-deformed commutation relation
by twisting the Hopf algebra of $igl(n,R)$
not by twisting the \poin Hopf algebra.
To understand why it is difficult to obtain
time-like $\kappa$-deformed commutation relation
by twisting the \poin Hopf algebra, $\cU(\cP)$,
it is instuctive to try a element
$\cF\in\cU(\cP)\otimes \cU(\cP)$ as a twist element.

For $\cF = 1+r+O(r^2)$, $r=r_1\otimes r_2$
for all $ r_1,r_2\in \cU(\cP)$
\begin{eqnarray}
\label{ncproduct}
x^\mu *  x^\nu &=&
\cdot \left[\cF
\triangleright (x^\mu \otimes x^\nu) \right]\nn\\
&=& x^\mu \cdot x^\nu+r_1(x^\mu)\cdot r_2(x^\nu)+\cdots
\end{eqnarray}

From the action of $r_1,r_2=(P_\rho, L_{\mu\nu})$,
where
$P_\rho\rightarrow -i \partial_\rho,~
L_{\mu\nu}\rightarrow -i (x_\mu\partial_\nu- x_\nu\partial_\mu)$, to the coordinates,
we infer the ansatz of the classical r-matrix
in order to obtain the first order relation of
the $\kappa$-deformed commutation relation,
Eq.~(\ref{nc2}), as

\begin{eqnarray}
r= r^{\rho\mu\nu}P_\rho\wedge L_{\mu\nu},
\label{rmatrix}
\end{eqnarray}
where $ r^{\rho\mu\nu}$ is a constant.

In order to this element $\cF$ to be a twist,
the above classical r-matrix have to satisfy
'classical Yang-Baxter equation',
i.e. the relation,

\beqa
[[r,r]]=0
\eeqa

Actually one can show that the above form of classical r-matrix in
Eq.(\ref{rmatrix}) cannot satisfy classical Yang-Baxter equation, in
general, except in a very special combination of
$P_\rho, L_{\mu\nu}$, i.e. $[P_\rho, L_{\mu\nu}]=0$,
which is the crucial condition for the twist to satisfy a 2-cocycle
condition. For that special case,
Lukierski and Woronowicz give the
'abelian' twist element
from  classical r-matrix \cite{LukiWoro}.
Although there have been many studies on field theories
in the $\kappa$-Minkowski spacetime, the attempts to
twist the Poincare group
to get the $\kappa$-deformed coordinate spacetime have succeeded
only in the light-cone $\kappa$-deformation \cite{Lukierski2},
\cite{LukiWoro}.

Hence, in order to obtain a twist which gives
time-like $\kappa$-deformation
we need a bigger symmetry
algebra than the \poin algebra.
We have obtain the $\kappa$-deformed non-commutativity
at the cost of bigger symmetry.
Our choice is the Hopf algebra of affine group $IGL(n,R)$
and we have successfully twisted $\cU(igl(n,R))$
in two different ways corresponding to
the $\kappa$-deformed and the $\theta$-deformed
non-commutativity.

Incidentally,
since there are two abelian subalgebras in $igl(n,R)$,
we derive two the non-commutativity
corresponding to
the $\kappa$-deformed and the $\theta$-deformed case
in section \ref{igl}.
Is there a twist
which transform the $\theta$-noncommutativity
to the $\kappa$-noncommutativity?
Since the inverses of $\cFth$ and $\cFk$ also satisfy the counital 2-cocycle condition,
%twist elements that gives the inverse twist,
we may think of the maps
$\cFth\circ\cFk^{-1}$  and $\cFk\circ\cFth^{-1}$ as mapping between the two different non-commutative spaces as
shown in  the following figure:

\begin{center}
\mbox{\large
\xymatrix{
\;&\mathcal{U}\mathrm{({\bf g})} \ar[ldd]_{\mathcal{F}_\theta} \ar[rdd]^{\mathcal{F}_\kappa} & \;\\
%\\
\\
\mathcal{U}_\theta\mathrm{({\bf g})} \ar[rr]_{\mathcal{F}_\kappa \circ \mathcal{F}^{-1}_\theta} & & \mathcal{U}_\kappa\mathrm{({\bf g})}~.
}
}
\end{center}

However, they ($\cFth\circ\cFk^{-1}$ and $\cFk\circ\cFth^{-1}$ ) do not satisfy
the counital 2-cocycle condition in general. Hence the composition map of them can not be a twist.
%Thus the family of twists do not form a group in general, so
We can not regard this relation between the two twists as a kind of
symmetry (while it can be a kind of symmetry in the context of
quasi-Hopf algebra).

The algebra $igl(n,R)$ we twisted in this paper can be
interpreted as a physical symmetry algebra in two ways.

One interpretation is to regard the $GL(4)$ as the generalization
of $O(1,3)$ \cite{Percacci, Smolin, Floreanini}.
In those works, they consider the generalization of the
Einstein-Hilbert action.
That is, from the theory of the local symmetry group
\begin{eqnarray}
\label{Diff} {\it Diff(M)}\times O(1,3),
\eeqa
to the theory of local symmetry group
\beqa
{\it Diff(M)}\times GL(4).
\end{eqnarray}
The present
method will be useful
when these kinds of non-commutativity
are applied to those theories with translational symmetry.

Another possible interpretation
 is to regard $igl(n,R)$ as the algebra of a subgroup of diffeomorphism group.
Since the Minkowski spacetime is a solution of Einstein-Hilbert
action, we obtain by twist the $\kappa$-Minkowski spacetime
and the canonical non-commutative spacetime using $\cFk$ and $\cFth$, respectively.
We may summarize this relations as
\begin{eqnarray}
\label{Diff}
\mbox{local symmetry group} &\backsimeq& {\it Diff(M)}\times O(1,3)\nn\\
&\backsimeq&
...\times \left(GL(4,R)\times O(1,3)\right)\ltimes T.
\end{eqnarray}
In this sense, we are in the same direction as in the work of Aschieri
et.al. \cite{Aschieri, Aschieri2}.
From the above relation,
we realize that the canonical twist in this interpretation has different origin
from the twists in earlier studies
\cite{chaichian},\cite{Wess}. We twist the subgroup  of the {\it Diff(M)},
while in earlier studies (for example, Chaichian et.al.~\cite{chaichian})
one twists the Hopf algebra, $o(1,3)$,
 i.e., the algebra of the Poincar\'{e} group,
\begin{equation}
\label{poincare}
P \backsimeq O(1,3)\ltimes T,
\end{equation}
or more general symmetry group (for example, conformal symmetry by Matlock \cite{Matlock}, etc).
The twisted $\kappa$-deformation gives the same coordinate non-commutativity,
but is distinguished from the $\kappa$-deformed Minkowski algebra in that
the co-algebra structures are different. 
%We note that the generalization of the $\kappa$-deformed twisting process in this paper
%to the whole diffeomorphism symmetry as in Aschieri et.al.¡¯s work
%\cite{Aschieri,Aschieri2} can also be possible.

%The twists discussed in this paper suffer no difficulties when it is applied to the quantum
%field theory. The irreducible representations of Poincar\'{e} group does
%not change because we twist the algebra of different symmetry ({\it Diff(M)} in our
%case) in principle. We can use the same Casimir operator as those in the corresponding commutative theory.
%The only chance of getting something new is to
%consider the twist inducing kappa-deformation. Taking into account
%the research profile of PLB an extension of the discussion presented
%in Sect.IV containing more concrete application of the results of the
%paper to twisted deformations of space-time diffeomorphisms is very
%desirable.

The blackholes in $\kappa$-deformed spacetime can be regarded as an example of the application of the result of this paper. Since field theories in  $\kappa$-deformed spacetime are the same as field theories with a $\kappa$-moyal product as in Eq.(\ref{kmoyal}) in commutative spacetime,
the deformed Einstein equations will be the ones in which the products between the metric and its derivatives are changed to the $\kappa$-moyal product. Among the solutions of the Einstein equations in commutative spacetime, a static solution is also a solution in a $\kappa$-deformed spacetime. The Einstein eqations are turned into the form:
\beqa
\label{dEinstein}
R_{\mu\nu}(g(x))=0 ~~\rightarrow~~  R_{\mu\nu}^\star(g(x))=0.
\eeqa
Since a static solution of the left hand side of Eq.(\ref{dEinstein}) has no time dependence, $\kappa$-moyal products in the right hand side are the same as normal products between them, it would automatically satisfy the deformed Einstein equations.
The  $\kappa$-deformed spacetime here denotes the module space of the twist induced algebra.
It should be distinguished from the $\kappa$-deformed spacetime which is a module space of the well-known $\kappa$-deformed algebra. 

Though a static solution is also a solution in $\kappa$-deformed spacetime, it has different dynamics. 
The time dependance changes the dynamics through $\kappa$-moyal products. That is, the perturbation equation of the static solution will be different:
\beqa
\label{dperturb}
\delta R^\star_{\mu\nu}(g(x))=0  ~~\ncong~~ \delta R_{\mu\nu}(g(x))=0.
\eeqa
Hence to tell the stability of the static solution in $\kappa$-deformed spacetime, careful analysis of the deformed perturbation equation Eq.(\ref{dperturb}) is expected. The stability analysis of these static solutions in $\kappa$-deformed spacetime is under investigation.

\begin{acknowledgments}
This work is supported in part by Korea Science and Engineering
Foundation Grant No. R01-2004-000-10526-0, and by the Korea Research
Foundation Grant funded by Korea Government(MOEHRD, Basic Research
Promotion Fund)(KRF-2005-075-C00009; H.-C.K.)
\end{acknowledgments}

\vspace{4cm}


\vspace{3cm}

\begin{thebibliography}{99}

\bibitem{Seiberg}
N. Seiberg, L. Susskind, and N. Toumbar, J. High Energy Phys. \textbf{06}, 044 (2000).

\bibitem{Gaume}
L. $\acute{\text{A}}$lvarez-Gaum$\acute{\text{e}}$, and J. L. F. Barbon. Int. J. Mod. Phys. {\bf A} \textbf{16}, 1123 (2001).

\bibitem{chaichian0}
M. Chaichian, K. Nishijima, and A. Tureanu, Phys. Lett. {\bf B} \textbf{568}, 146 (2003).

\bibitem{Gomis}
J. Gomes, and T. Mehen, Nuclear Physics {\bf B} \textbf{591}, 265 (2000).

\bibitem{Bahns}
D. Bahns, S. Doplicher, K. Fredenhagen, and G. Piacitelli, Phys. Lett. {\bf B} \textbf{533}, 178 (2002).

\bibitem{Lukierski}
J. Lukierski, A. Nowicki,
H. Ruegg and V. N. Tolstory, Phys. Lett. \textbf{B 264}, 331 (1991).

\bibitem{Doplicher}
S. Doplicher, K. Fredenhagen, and J. E. Roberts, Comm. Math. Phys. \textbf{172}, 187 (1995).

\bibitem{Weyl}
H. J. Groenewold, Physica \textbf{12}, 405 (1946); J. E. Moyal,
Proc. Cambridge Phil. Soc. \textbf{45}, 99 (1949); H. Weyl,
\textit{Quantum mechanics and group theory}, Z. Phys. \textbf{46}, 1 (1927).

\bibitem{Bahns1}
D. Bahns, Fortsch. Phys. \textbf{51}, 658 (2003).

\bibitem{Bahns2}
D. Bahns, Fortsch. Phys. \textbf{52}, 458 (2004).

\bibitem{BahnDo}
D. Bahns, S. Doplicher, K. Fredenhagen, and G. Piacitelli, Comm. Math. Phys. \textbf{237}, 221 (2003).

\bibitem{Liao}
Y. Liao, and K. Sibold, Eur. Phys. J. {\bf C} \textbf{25}, 469 (2002).

\bibitem{Yee}
C. Rim, and J. H. Yee, Phys. Lett. {\bf B} \textbf{574}, 111 (2003).

\bibitem{Drinfeld}
V. G. Drinfeld, Soviet Math. Dock. \textbf{32}, 254 (1985).

\bibitem{Jimbo}
M. Jimbo, Lett. Math. Phys. \textbf{10}, 63 (1985).

\bibitem{Baez}
J. C. Baez, and A. D. Lauda, {\it A History of n-Categorical Physics}, Draft Version (2006).

\bibitem{Oeckl}
R. Oeckl, Nucl. Phys. \textbf{B 581}, 559 (2000).

\bibitem{chaichian}
M. Chaichian, P. P. Kulish, K. Nishijima, and A.Tureanu, Phys. Lett. \textbf{B 604}, 98(2004).

\bibitem{Wess}
J. Wess, hep-th/0408080.

\bibitem{Gonera}
C. Gonera, P. Kosinski, P. Maslanka, and S. Giller Phys.Lett. \textbf{B 622}, 192 (2005).

\bibitem{Matlock}
P. Matlock, Phys. Rev. {\bf D} \textbf{71}, 126007 (2005).

\bibitem{Lizzi}
F. Lizzi, S. Vaidya, and P. Vitale, Phys. Rev. \textbf{D 73}, 125020 (2006).

\bibitem{Choonkyu}
R. Banerjee, C. Lee, and S. Siwach, hep-th/0511205.

\bibitem{Saemann}
M. Ihl and C. Saemann, JHEP, {\bf 0601}, 065 (2006).

\bibitem{Kobayashi}
Y. Kobayashi, and S. Sasaki, Int. J. Mod. Phys. {\bf A 20}, 7175 (2005).

\bibitem{Sunandan}
B. Chakraborty, S. Gangopadhyay, A. G. Hazra, and F. G. Scholtz, J. Phys. \textbf{A 39}, 9557 (2006).

\bibitem{Banerjee}
R. Banerjee, Eur. Phys. J. \textbf{C 47}, 541 (2006).

\bibitem{Vassil}
D. V. Vassilevich, Mod. Phys. Lett. \textbf{A 21}, 1279 (2006).

\bibitem{Wess0}
P. Aschieri, M. Dimitrievic, F. Meyer, S. Schraml and J.Wess, hep-th/0603024.

\bibitem{Archil}
A. Kobakhidze, hepth/0603132.

\bibitem{Aschieri}
P. Aschieri, C. Blohmann, M. Dimitrijevi$\acute{\text{c}}$, F.Meyer,
P. Schupp, and J. Wess, Class. Quant. Grav. \textbf{22}, 3511 (2005).

\bibitem{Aschieri2}
P. Aschieri, M. Dimitrijevic, F. Meyer, and J. Wess, Class. Quant. Grav. \textbf{23}, 1883 (2006).

\bibitem{Saemann2}
S. Kurkcuoglu and C. Saemann, hep-th/0606197.

\bibitem{chaichian2}
M. Chaichian, P. Presnajder, and  A. Tureanu, Phys. Rev. Lett.
\textbf{94}, 151602 (2005).

\bibitem{Bala}
A. P. Balachandran, G. Mangano, A. Pinzul and S. Vaidya, Int. J. Mod. Phys. \textbf{A 21}, 3111 (2006).

\bibitem{Zahn}
J. Zahn, Phys. Rev. \textbf{D 73}, 105005 (2006).

\bibitem{Bu}
J. G. Bu, H. C. Kim, Y. Lee, C. H. Vac, and J. H. Yee, Phys. Rev.  \textbf{D 73}, 125001 (2006).

\bibitem{Lukierski2}
J. Lukierski, V. Lyakhovsky, and M. Mozrzymas, Phys. Lett. \textbf{B 538}, 375 (2002).

\bibitem{LukiWoro}
J. Lukierski, M. Woronowicz, hep-th/0508083.

\bibitem{LukiWoro2}
J. Lukierski, and M. Woronowicz, hep-th/0512046.

\bibitem{Majid}
S. Majid, \textit{Foundations of Quantum Group Theory}, Cambridge University Press, (1995).

\bibitem{Percacci}
R. Percacci, Phys. Lett. \textbf{B 144}, 37 (1984);
R. Percacci \textit{Geomety of Nonlinear Field Theories}, Singapore: World
Scientific, 1986;


\bibitem{Smolin}
L. Smolin, Nucl. Phys. \textbf{B 132}, 138 (1978).

\bibitem{Floreanini}
R. Floreanini, and R. Percacci, Class. Quantum Grav. \textbf{7}, 975 (1990).

\bibitem{chaichian3}
M. Chaichian, A. P. Demichev, and N. F. Nelipa, Comm. Math. Phys. \textbf{90}, 353 (1983).



\end{thebibliography}
\end{document}